\documentclass[aps,prd,amsmath,amssymb,showpacs]{revtex4}

\usepackage{graphicx}
\begin{document}

\title{Top Quark Spin Polarization in $e \gamma$ Collision}

\author{B. \c{S}ahin}
\email[]{dilec@science.ankara.edu.tr}
\affiliation{Department of Physics, Faculty of Sciences,
Ankara University, 06100 Tandogan, Ankara, Turkey}

\author{\.{I}. \c{S}ahin}
\email[]{isahin@science.ankara.edu.tr} \affiliation{Department of
Physics, Faculty of Sciences, Ankara University, 06100 Tandogan,
Ankara, Turkey}

\begin{abstract}
We investigate the degree of spin polarization of single top
quarks in the $e\gamma$ collision via the process $e^{+} \gamma
\to t \bar{b} \bar{\nu_{e}}$ with center of mass energies
$\sqrt{s}$=0.5, 1 and 1.5 TeV of the parental linear $e^{+}e^{-}$
collider. Dominant spin fractions and spin asymmetries for the
various top quark spin bases are investigated. We show that
$e^{+}$-beam direction is the favorite top quark spin
decomposition axis. It is found to be comparable with the ones in
$pp$ and $ep$ collisions.

\end{abstract}


\maketitle
\begin{quote}
Preprint of this article submitted for consideration in Modern
Physics Letters A (2007).
\end{quote}
\section{Introduction}

The top quark is the heaviest fermion in the Standard Model. Its
mass is at the electroweak symmetry-breaking scale. Due to its
large mass its weak decay time is much shorter than the typical
time for the strong interactions to affect its spin \cite{bigi}.
Therefore the information on its polarization, is not disturbed by
hadronization effects but transferred to the decay products.
Within the standard model, the dominant decay of the top quark is
$t \to W^{+} b$. The angular distributions of the top quark decay
products are determined by the momentum and spin state of the top
quark itself. It is believed that top quark physics will be
crucial to understand the structure of Standard Model and
existence of new physics beyond. Detailed discussions on top quark
spin polarization properties will contribute future researches on
this subject.

In this work we investigate top quark spin polarization along the
direction of various spin bases for the single production in
$e^{+} \gamma$ collision via the process $e^{+} \gamma \to t
\bar{b} \bar{\nu_{e}}$. The research and development on linear
$e^{+}e^{-}$ colliders have been progressing and the physics
potential of these future machines is under study. After linear
colliders are constructed its operating modes of $e\gamma$ and
$\gamma\gamma$ are expected to be designed \cite{akerlof,barklow}.
Real gamma beam is obtained through Compton backscattering of
laser light off linear electron beam where most of the photons are
produced at the high energy region. The luminosities for $e\gamma$
and $\gamma\gamma$ collisions turn out to be of the same order as
the one for $e^{+}e^{-}$ \cite{Ginzburg}, so the cross sections
for photoproduction processes with real photons are considerably
larger than virtual photon case. In our calculations we consider
three different center of mass energies $\sqrt{s}$=0.5, 1 and 1.5
TeV of the parental linear $e^{+}e^{-}$ collider.

There are many detailed discussions in the literature for single
top quark production and spin correlations in $pp$, $p\bar{p}$ and
$ep$ collisions \cite{parke,mahlon,atag}. At $pp$ and $p\bar{p}$
colliders top quarks are produced $t \bar{t}$ and single t
production modes. In double production case, spin up top quark and
spin down anti-top quark are more likely produced and there are
observable angular correlations among the decay products. $t
\bar{t}$ decay products in the final states give larger statistics
than the single production case. On the other hand, final state in
the single top case is relatively simple and may compensate for
the smaller statistics \cite{mahlon}. In $ep$ collisions, top
quark decay products in the final states are dominated by the
single top production due to absence of the $t \bar{t}$
production. Since single top production contains an electroweak
process with produced top quarks coupled to a W boson, decay
products of the final top quark gives a high angular correlation
\cite{atag}. This is also the case in the $e^{+} \gamma$
collision. Moreover linear $e^{+}e^{-}$ collider or its $e\gamma$
mode provide a clean environment to study polarization phenomena
of top quarks.

\section{Cross Sections of Polarized Top Quarks}

In $e^{+}e^{-}$ linear colliders a hard photon beam can be produced
by Compton backscattering of laser light off linear electron or
positron beam. We consider the case in which the photon beam is
obtained by Compton backscattering off linear electron beam and the
positron beam directly takes part in the subprocess. The spectrum of
the backscattered photons is given by \cite{Ginzburg}.

\begin{eqnarray}
f_{\gamma/e}(y)={{1}\over{g(\zeta)}}[1-y+{{1}\over{1-y}}
-{{4y}\over{\zeta(1-y)}}+{{4y^{2}}\over {\zeta^{2}(1-y)^{2}}}]
\end{eqnarray}

where,

\begin{eqnarray}
g(\zeta)=&&(1-{{4}\over{\zeta}}
-{{8}\over{\zeta^{2}}})\ln{(\zeta+1)}
+{{1}\over{2}}+{{8}\over{\zeta}}-{{1}\over{2(\zeta+1)^{2}}}
\end{eqnarray}

with $\zeta=4E_{e}E_{0}/M_{e}^{2}$. $E_{0}$ is the energy of
initial laser photon and $E_{e}$ is the energy of initial electron
beam before Compton backscattering. $y$ is the fraction which
represents the ratio between the scattered photon and initial
electron energy for the backscattered photons moving along the
initial electron direction. Maximum value of $y$ reaches 0.83 when
$\zeta=4.8$ in which the backscattered photon energy is maximized
without spoiling the luminosity. The integrated cross section over
the backscattered photon spectrum is given by:

\begin{eqnarray}
\sigma(s)=\int_{y_{min}}^{0.83}
f_{\gamma/e}(y)\hat{\sigma}(\hat{s}) dy
\end{eqnarray}

where $y_{min}=\frac{m_{t}^{2}}{s}$ and $\hat{s}$ is the square of
the center of mass energy of the subprocess $e^{+} \gamma \to t
\bar{b} \bar{\nu_{e}}$. $\hat{s}$ is related to $s$, the square of
the center of mass energy of  $e^{+}e^{-}$ by $\hat{s}=ys$.

The single production of top quark via the process $e^{+} \gamma
\to t \bar{b} \bar{\nu_{e}}$ is described by four tree level
diagrams. Each of the diagrams contains Wtb vertex and due to its
V-A structure, produced top quarks are highly polarized. Because
of its large mass the helicity of a top quark is frame dependent
and changes under a boost from one frame to an other. The helicity
and chirality states do not coincide with each other and there is
no reason to believe that the helicity basis will give the best
description of the spin of top quarks. So it is reasonable to
study other spin bases better than helicity for top quark spin.

The spin four-vector of a top quark is defined by

\begin{eqnarray}
s_{t}^{\mu}=(\frac{\vec{p}_{t}\cdot \vec{s^{\prime}}}{m_{t}} \,,\,
\vec{s^{\prime}}+\frac{\vec{p}_{t}\cdot \vec{s^{\prime}}
}{m_{t}(E_{t}+m_{t})}\vec{p}_{t})
\end{eqnarray}

where $(s_{t}^{\mu})_{RF}=(0,\vec{{s}^{\prime}})$ in the top quark
rest frame. The laboratory frame is the $e^{+}e^{-}$ center of
mass system where the cross section is performed. We consider four
different top spin direction in the laboratory frame; the incoming
positron beam, photon beam directions and outgoing $\bar{b}$
direction and also the helicity basis. $\vec{{s}^{\prime}}$ should
be obtained by a Lorentz boost from laboratory frame:

\begin{eqnarray}
  \vec{{s}^{\prime}}=\lambda \frac{\vec{{p}^{\star}}}
  {|\vec{{p}^{\star}}|} ,\,\,\,\,\, \lambda=\pm 1.\nonumber\\
\vec{{p}^{\star}}=\vec{p}+\frac{\gamma-1}{\beta^{2}}
(\vec{\beta}\cdot \vec{p})\vec{\beta}
 -E\gamma \vec{\beta}
\end{eqnarray}

Here $\vec{p}$  is the momentum of the particle moves along the
top spin direction in the laboratory frame and $\vec{{p}^{\star}}$
is the momentum observed in the rest frame of the top quark.

In Table I-III polarized cross sections, dominant spin fractions
and spin asymmetries are given for the various top quark spin
bases. It is shown from these tables that $e^{+}$-beam direction
gives the highest degree of polarization; $94\%$ at $\sqrt{s}=0.5$
TeV, $88\%$ at $\sqrt{s}=1$ TeV and $86\%$ at $\sqrt{s}=1.5$ TeV.
Its energy dependence is significant. In relatively low energies,
helicity will not give the best description of the spin of a
massive fermion like top quark. Therefore other spin bases may
give a better description of top quark spin. In our case favorite
spin basis is the $e^{+}$-beam spin decomposition axis. When
energy increases top quark gradually becomes ultrarelativistic and
spin fraction in the helicity basis grows. This behaviour is shown
from Table I-III. Spin fraction for the helicity basis increases
with energy; $60\%(L)$ at $\sqrt{s}=0.5$ TeV, $66\%(L)$ at
$\sqrt{s}=1$ TeV and $69\%(L)$ at $\sqrt{s}=1.5$ TeV. At the
energy region we have considered speed of the top quarks are not
ultrarelativistic therefore helicity is not the favorite spin
basis.

It is straightforward to obtain similar results for anti top quarks
with the process $e^{-} \gamma \to \bar{t} b \nu_{e}$. In this case
favorite spin basis is the $e^{-}$-beam spin decomposition axis.
Same results at Table I-III for the spin fractions are obtained with
the interchange of the bases; $e^{+}$-beam $\longleftrightarrow$
$e^{-}$-beam , $\bar{b}$-beam $\longleftrightarrow$ $b$-beam. But
the spin orientations should be reversed; spin up
$\longleftrightarrow$ spin down, L $\longleftrightarrow$ R.

Another useful quantity about spin-induced angular correlations is
the spin asymmetry

\begin{eqnarray}
 A_{\uparrow\downarrow}=\frac{N_{\uparrow}-N_{\downarrow}}{N_{\uparrow}+N_{\downarrow}}
\end{eqnarray}

Here subscript up arrow $\uparrow$ (down arrow $\downarrow$) stands
for spin up $\lambda=+1$ (spin down $\lambda=-1$) and $N$ represents
number of event for the corresponding spin. The angular distribution
of the top quark decay involves correlations between top decay
products and top quark spin:

\begin{eqnarray}
 \frac{1}{\Gamma_{T}}\frac{d\Gamma}{dcos\theta}=\frac{1}{2}(1+A_{\uparrow\downarrow}\alpha cos\theta)
\end{eqnarray}

Here the dominant decay chain of the top quark in the standard
model $t \to W^{+}b(W^{+} \to l^{+}\nu,\bar{d}u)$ is considered.
$\theta$ is defined as the angle between top quark decay products
and the top quark spin quantization axis in the rest frame of the
top quark. $\alpha$ is the correlation coefficient and $\alpha=1$
for $l$ or $\bar{d}$ which leads to the strongest correlation. One
can see from Table I that $e^{+}$-beam basis improves the
asymmetry a factor of 4.35 when compared to helicity basis at
$\sqrt{s}=0.5$ TeV. From Table II and III we see that this factor
takes the value of 2.38 and 1.85 respectively.

In order to get an idea about the influence of spin polarization
on transverse momentum $P_{t}^{T}$ distributions of singly
produced top quarks we plot Fig I-III. Similar features in the
tables are reflected in the figures also; deviations of the
$P_{t}^{T}$ distributions from the unpolarized (total) is the
largest for helicity basis.

In our calculations phase space integrations have been performed
by GRACE \cite{grace} which uses a Monte Carlo routine.

\section{Conclusions}

We have shown that in the energy region $\sqrt{s}=0.5-1.5$ TeV
$e^{+}$-beam direction provides a high degree of polarization.
Considerable improvements have been obtained in the spin fractions
and asymmetries with respect to helicity basis. Therefore a
detailed analysis of top spin polarization and studying various
spin bases in the $e^{+} \gamma$ collision is important and may
contribute future researches.

Linear $e^{+}e^{-}$ collider and its $e\gamma$ mode provide a clean
environment and the experimental clearness is an advantage of
$e\gamma$ collisions with respect to $pp$, $p\bar{p}$ and $ep$
collisions. At hadron colliders the interacting particles are not
the beam particles themselves. The reactions are initiated by one of
many partons present in the incident hadron. So the energy and
quantum state of the initial state are not fixed. On the other hand
lepton colliders tend have much cleaner beams and lower backgrounds.




\begin{figure}
\includegraphics{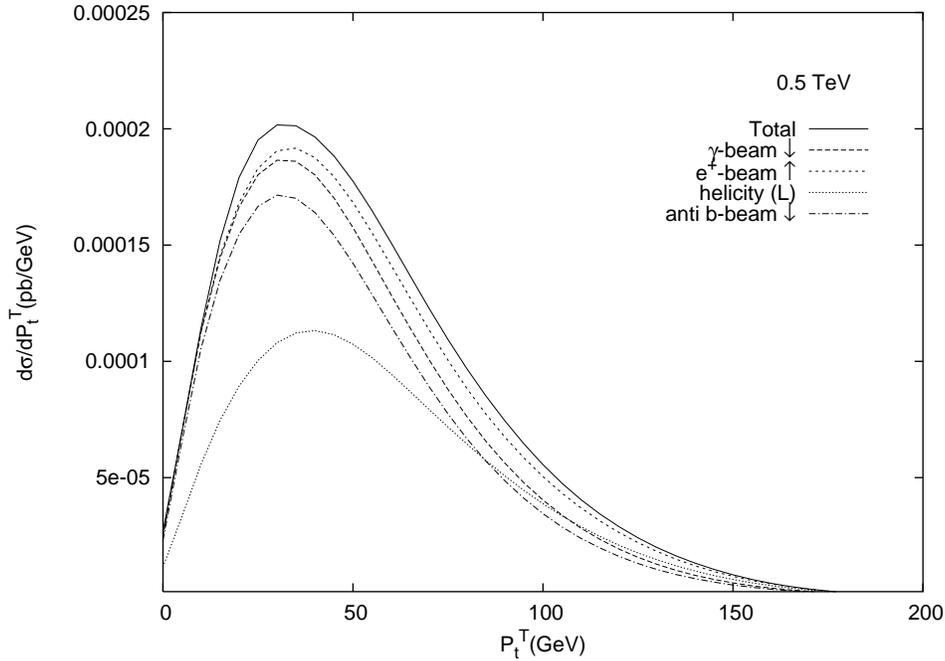}
\caption{Transverse momentum $P_{t}^{T}$ distributions of the
singly produced top quarks via process $e^{+} \gamma \to t \bar{b}
\bar{\nu_{e}}$ at center of mass energy $\sqrt{s}=0.5$ TeV of the
parental linear $e^{+}e^{-}$ collider. Dominant spin bases
$\gamma$-beam down, $e^{+}$-beam up, helicity left, $\bar{b}$-beam
down and unpolarized (Total) case are drawn. \label{fig1}}
\end{figure}

\begin{figure}
\includegraphics{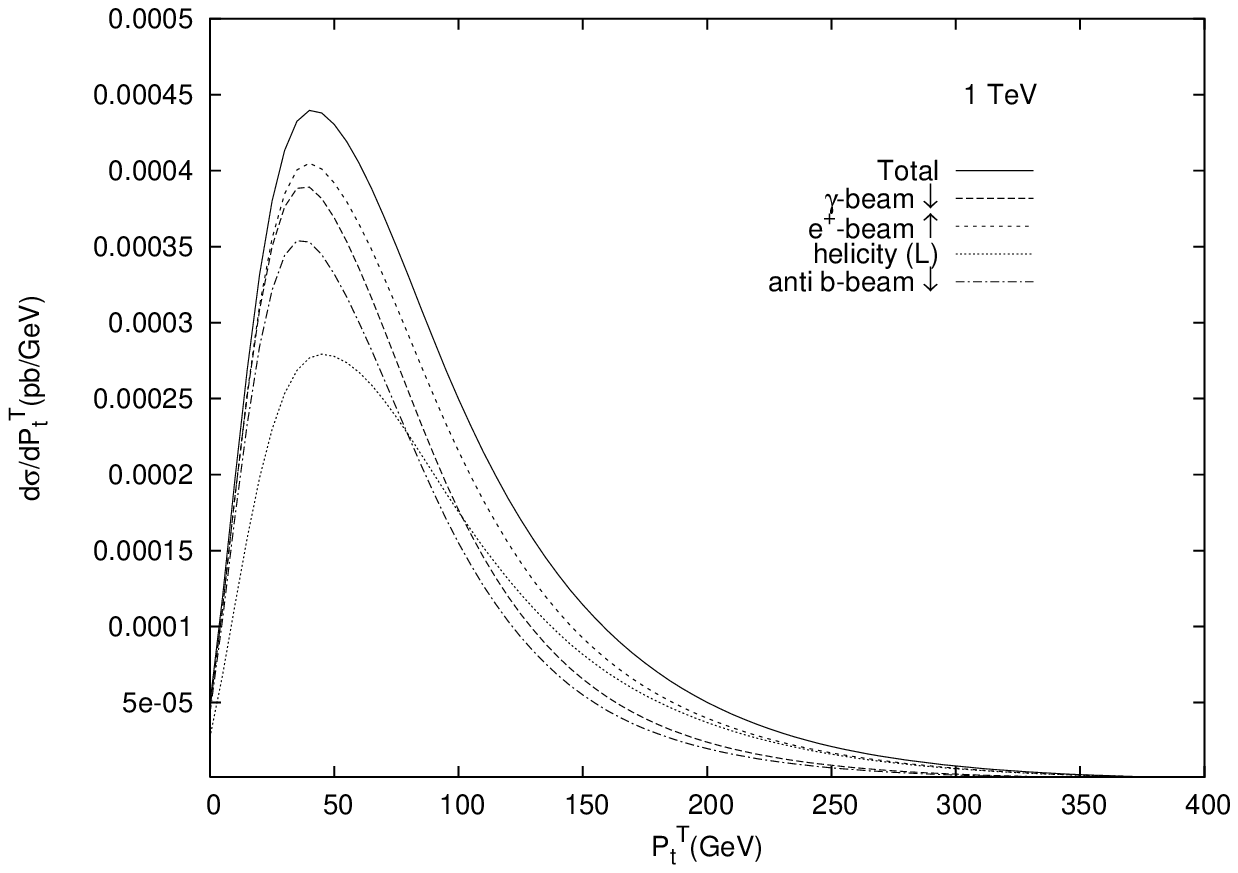}
\caption{The same as FIG. 1 but for $\sqrt{s}=1$ TeV.\label{fig2}}
\end{figure}

\begin{figure}
\includegraphics{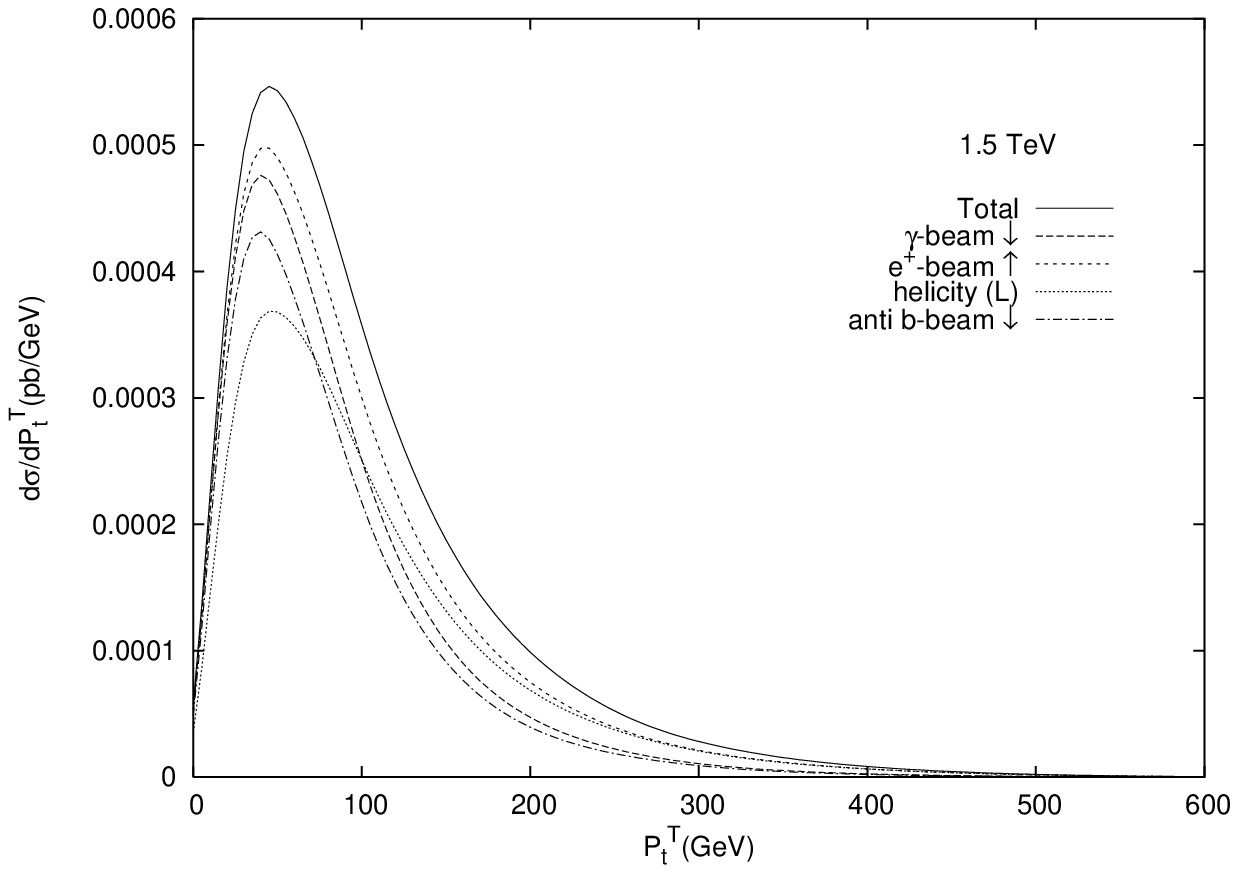}
\caption{The same as FIG. 2 but for $\sqrt{s}=1.5$
TeV.\label{fig3}}
\end{figure}


\begin{table}
\caption{Polarized cross sections, dominant spin fractions and
spin asymmetries for the various top quark spin bases in the
production of single top process $e^{+} \gamma \to t \bar{b}
\bar{\nu_{e}}$ at center of mass energy $\sqrt{s}=0.5$ TeV of the
parental linear $e^{+}e^{-}$ collider. \label{tab1}}
\begin{ruledtabular}
\begin{tabular}{cccc}
basis &polarized cross sections (pb) &spin fractions & $\frac{N_{\uparrow}-N_{\downarrow}}{N_{\uparrow}+N_{\downarrow}}$\\
\hline
$e^{+}$-beam & 0.014& 94\%$\uparrow$ &0.87 \\
$\gamma$-beam & 0.013& 86\%$\downarrow$ &-0.73 \\
$\bar{b}$-beam & 0.012& 78\%$\downarrow$ &-0.60\\
helicity & 0.009& 60\%(L) &-0.20  \\
\end{tabular}
\end{ruledtabular}
\end{table}

\begin{table}
\caption{The same as table I but for $\sqrt{s}=1$ TeV.
\label{tab2}}
\begin{ruledtabular}
\begin{tabular}{cccc}
basis &polarized cross sections (pb) &spin fractions & $\frac{N_{\uparrow}-N_{\downarrow}}{N_{\uparrow}+N_{\downarrow}}$\\
\hline
$e^{+}$-beam &0.044 & 88\%$\uparrow$ &0.76  \\
$\gamma$-beam &0.038 & 76\%$\downarrow$ &-0.52\\
$\bar{b}$-beam &0.034 & 68\%$\downarrow$ &-0.36 \\
helicity &0.033 &66\%(L) &-0.32  \\
\end{tabular}
\end{ruledtabular}
\end{table}

\begin{table}
\caption{The same as table II but for $\sqrt{s}=1.5$ TeV.
\label{tab3}}
\begin{ruledtabular}
\begin{tabular}{cccc}
basis &polarized cross sections (pb) &spin fractions & $\frac{N_{\uparrow}-N_{\downarrow}}{N_{\uparrow}+N_{\downarrow}}$\\
\hline
$e^{+}$-beam &0.062 & 86\%$\uparrow$ &0.72  \\
$\gamma$-beam &0.052 & 72\%$\downarrow$ &-0.44 \\
$\bar{b}$-beam &0.047 & 65\%$\downarrow$ &-0.31\\
helicity &0.050 & 69\%(L) &-0.39  \\
\end{tabular}
\end{ruledtabular}
\end{table}

\end{document}